\documentclass[11pt]{article}

\usepackage{moreverb,url,a4wide}
\usepackage{graphicx}
\graphicspath{{figures/}}
\usepackage{hyperref}
\usepackage[round]{natbib}

\begin{document}

\graphicspath{{figures/}}

\title{An Application of Mosaic Diagrams to the Visualization of Set Relationships}

\author{Saturnino Luz \\
  Usher Institute of Population Health Sciences and Informatics\\
  The University of Edinburgh, Scotland, UK \and
  Masood Masoodian \\
  School of Arts, Design and Architecture\\
    Aalto University, Finland\\ 
    masood.masoodian@aalto.fi}

\maketitle

\begin{abstract}
We present an application of mosaic diagrams to the visualisation of set relations. Venn and Euler diagrams are the best known visual representations of sets and their relationships (intersections, containment or subsets, exclusion or disjointness). In recent years, alternative forms of visualisation have been proposed. Among them, linear diagrams have been shown to compare favourably to Venn and Euler diagrams, in supporting non-interactive assessment of set relationships. Recent studies that compared several variants of linear diagrams have demonstrated that users perform best at tasks involving identification of intersections, disjointness and subsets when using a horizontally drawn linear diagram with thin lines representing sets, and employing vertical lines as guide lines. The essential visual task the user needs to perform in order to  interpret this kind of diagram is vertical alignment of parallel  lines and detection of overlaps. Space-filling mosaic diagrams which  support this same visual task have been used in other applications,  such as the visualization of schedules of activities, where they  have been shown to be superior to linear Gantt charts. In this  paper, we present an application of mosaic diagrams for  visualization of set relationships, and compare it to linear  diagrams in terms of accuracy, time-to-answer, and subjective  ratings of perceived task difficulty. The study participants exhibited similar performance on both visualisations,  suggesting that mosaic diagrams are a good alternative to Venn and Euler diagrams, and that the choice  between linear diagrams and mosaics may be solely guided by visual design considerations. 
\end{abstract}

{\bf keywords:} Set visualization, set relationships, linear diagrams,
  mosaic diagrams, space-filling visualizations, visual design.

\section{Introduction}
\label{sec:introduction}

The study of sets and their relationships is fundamental to the
disciplines of mathematics, logic, and computer science. Visual
representations of relationships among sets --- intersection,
containment, and exclusion (disjoint sets) --- have been used for
centuries. However, the development of interactive visualizations and
tools in recent years has gained a new impetus due to the wide range
of applications that these tools find in a variety of areas, including
the analysis of healthcare and population data, representation of
relationships in social networks, and the study of consumer purchasing
patterns, to name a few.

Visual representation of sets and their relationships is most commonly
done through Venn and Euler diagrams \citep{bib:Baron1969}. These types
of diagrams, however, have well known
limitations \citep{bib:Rodgers2014,bib:Gottfried2014,bib:Gottfried2015}.
They generally do not scale well beyond a small number of sets, and
present usability problems. Automatic drawing of Venn and Euler
diagrams is also problematic \citep{bib:RicheAndDwyer2010,bib:FlowerEtal2014,bib:Simonetto2016}. In
response to these limitations, alternative set visualization
techniques have been proposed.

Linear diagrams \citep{bib:Gottfried2014}, which are of particular
interest here, have been shown to compare favourably to Venn and Euler
diagrams in terms of task completion time and the number of errors
made by users \citep{bib:ChapmanEtal2014}, for example in tasks
involving syllogistic reasoning \citep{bib:SatoAndMineshima2012}.

In a recent paper \cite{bib:RodgersEtal2015} compared
several versions of linear diagrams produced by varying essential
properties of their corresponding {\em retinal} and {\em planar}
variables \citep{bib:Bertin67Semiologie}. Their study concluded that users
perform best at tasks involving identification of intersections,
disjointness (exclusion) and subsets (inclusion) among sets when using
a horizontally drawn linear chart with thin lines representing sets,
and vertical guide lines for aiding the detection of alignment across
the vertical axis. The essential visual task the user needs to perform
in order to interpret this kind of linear diagrams is a Vernier acuity
task, which basically requires vertical alignment of the beginning or
end of a horizontal line with those of other lines above or below.

In this paper, we present a study comparing linear diagrams with a
space-filling alternative visualization based on mosaic diagrams,
\citep{bib:LuzMasoodian2007} for representing set relationships as
examined by \cite{bib:RodgersEtal2015}.  The proposed mosaic diagrams
(as shown below) employ a space-filling algorithm whereby
intersections are denoted by shared areas (represented in different
colours), subset relations are denoted by area containment, and
exclusions are represented as uniformly (i.e. single) coloured areas.

The primary motivation for this study was the fact that mosaic
diagrams have previously been used in other applications (e.g.
visualization of task schedules) where they have been shown to be
superior to linear-style diagrams such as Gantt
charts \citep{bib:Gantt19}. The current study therefore investigates
whether this superiority of mosaic over linear diagrams also holds true
in the case of set visualization. To the best of our knowledge, this
study is the first to compare mosaic and linear diagrams in set
visualization tasks. As such, it aims specifically at comparing static
representations of set relations, and using two representations both
based on linear structure, albeit employing different instantiations
of retinal and planar 
variables (see further discussion below), rather than providing an
exhaustive comparison of set visualisation methods based on disparate
principles, or replicating previous comparisons. Thus it does not
compare mosaic or linear diagrams to other static representations such
as Euler diagrams, Venn diagrams or their modern variants described
below, as comparisons between linear diagrams and Euler and Venn
diagrams have been reported
elsewhere \citep{bib:Gottfried2015,bib:ChapmanEtal2014}. In particular, as regards
variants such as Bubble
Sets \citep{bib:CollinsEtal2009,bib:RicheAndDwyer2010} and
LineSets \citep{bib:AlperEtal2011}, these techniques are unlike linear
(and mosaic) diagrams, which display only abstract set relations, in
that they ``require the existence of embedded items'', as pointed out
by \cite{bib:RodgersEtal2015}. Similarly, this study does
not compare mosaics to the many interactive set visualization systems
proposed in the burgeoning literature on this topic. The reader is
referred to these works and to the literature review below for comparisons
of interactive systems in terms of their design
features \citep{bib:YalcinEtal2016} and task
taxonomies \citep{bib:AlsallakhEtal2015}. While empirical studies of
interactive versions of mosaic (and indeed linear diagrams) are of
great practical interest for future work, comparisons of this kind lie
beyond our scope here.

This paper contributes to the information visualization
literature by providing an analysis of time-on-task, accuracy and subjective
difficulty ratings for each of these two linearly-structured visualizations,
with essentially comparable forms of set representation. In addition to its empirical
findings, the paper also discusses the relative advantages and
disadvantages of mosaic and linear diagrams in terms of their design, including their
potential uses as compact overviews of sets, and their ability to
represent other properties (e.g. cardinality) of sets beyond the basic set
relationships investigated in the current study.

\section{Background}

\subsection{Set visualizations}
\label{sec:lit-review}

Set visualization is a common and increasingly important task.  Not
surprisingly, a wide range of set visualization techniques have been
proposed over the years. 
\cite{bib:AlsallakhEtal2015,bib:AlsallakhEtal2014} provide a
comprehensive review of set visualizations in their state-of-the-art
report. They classify set visualizations into six categories:

\begin{enumerate}
\item \emph{Euler and Venn diagrams:} As mentioned, these
  visualizations are the most common representations of sets, and a
  large number of variations have been designed to improve them. For
  surveys, see \cite{bib:Rodgers2014} and \cite{bib:RuskeyAndWeston2005}.

\item \emph{Overlays:} These techniques present set memberships as
  secondary information over other visualizations (e.g. spatial, or temporal) which
  provide the context for analysis. These include the popular 
  LineSets \citep{bib:AlperEtal2011}, Bubble Sets \citep{bib:CollinsEtal2009},
  Kelp diagrams \citep{bib:MeulemansEtal2013}, and TimeSets \citep{bib:NguyenEtal2016}.

\item \emph{Node-link diagrams:} These techniques represent
  relationships between sets and their members as edges of bipartite
  graphs whose nodes are the sets and elements. Node-link diagrams are
  considered to be easy to understand, and allow visual encoding of
  further information in representation of the nodes (i.e. each element or
  set). Node-link visualizations can also be combined with other
  representations such as matrix-based (e.g. OnSet
  \citep{bib:Sadana2014}), or aggregation-based (e.g. Radial Sets
  \citep{bib:AlsallakhEtal2013}) representations.

\item \emph{Matrix-based techniques:} These visualizations use the
  matrix representation to show sets or set members as elements of
  matrices.  Examples of this type of visualizations include UpSet \citep{bib:LexEtal2014},
   and OnSet \citep{bib:Sadana2014}.

\item \emph{Aggregation-based techniques:} Unlike some of the above
  mentioned techniques, aggregation-based visualizations do not aim to
  represent the relationships between individual elements of the sets
  involved.  Instead, set elements are aggregated into their
  respective sets, and only relationships between those sets are
  represented. As such, aggregation-based techniques are more suitable
  for representing relationships between sets with large number of
  elements, where it would be impractical to show all the
  relationships between those elements.  Examples of these techniques
  include AggreSet \citep{bib:YalcinEtal2016}, Radial Sets \citep{bib:AlsallakhEtal2013},
   and PowerSets \citep{bib:AlsallakhAndRen:2017}.

\item \emph{Other techniques:} There are also a range of other set
  visualization techniques, such as Scatter plots (e.g. scatter view
  and cluster view\citep{bib:AlsallakhEtal2015}), which represent set
  relationships using other visual methods than those described in the
  above categories. These include techniques such as bargrams, which
  resemble linear diagrams in some aspects but incorporate other
  extensions, such as set-valued
  attributes \citep{bib:WittenburgMaliziaEtAl12vsatp}, and can be
  categorised as frequency-based.
\end{enumerate}

More specifically, linear diagrams \citep{bib:Gottfried2014} 
fall into the category of aggregation-based techniques.
Linear diagrams have been shown to be more effective than region-based 
representation such as Euler and Venn diagrams \citep{bib:ChapmanEtal2014}, 
which tend to be more cluttered due to overlapping, coincident, and tangentially 
touching contours, as demonstrated in an empirical study \citep{bib:Gottfried2015}.

As mentioned earlier, \cite{bib:RodgersEtal2015} have
also conducted a series of studies which compared the effectiveness of
linear diagrams against Euler and Venn diagrams, as well as different
variations of linear diagrams themselves, for preforming tasks
requiring visualization of set relationships. These studies have shown
that linear diagrams are superior to Euler and Venn diagrams for
identification of set intersections, containment, and exclusions. They
have also led to a number of visual design principles for creating
more effective linear diagrams. These include: a) \emph{the use of a
  minimal number of line segments}, b) \emph{the use of guide lines
  where line overlaps start and end}, and c) \emph{the use of lines
  that are thin as opposed to thick bars} \citep{bib:RodgersEtal2015}.
The effectiveness of these principles was demonstrated through a final
study \citep{bib:RodgersEtal2015}, which we utilize in our own study,
presented in this paper.

\subsection{Mosaic diagrams}
\label{sec:mosaic}

\begin{figure}[!t]
\centering
\includegraphics[width=0.45\linewidth]{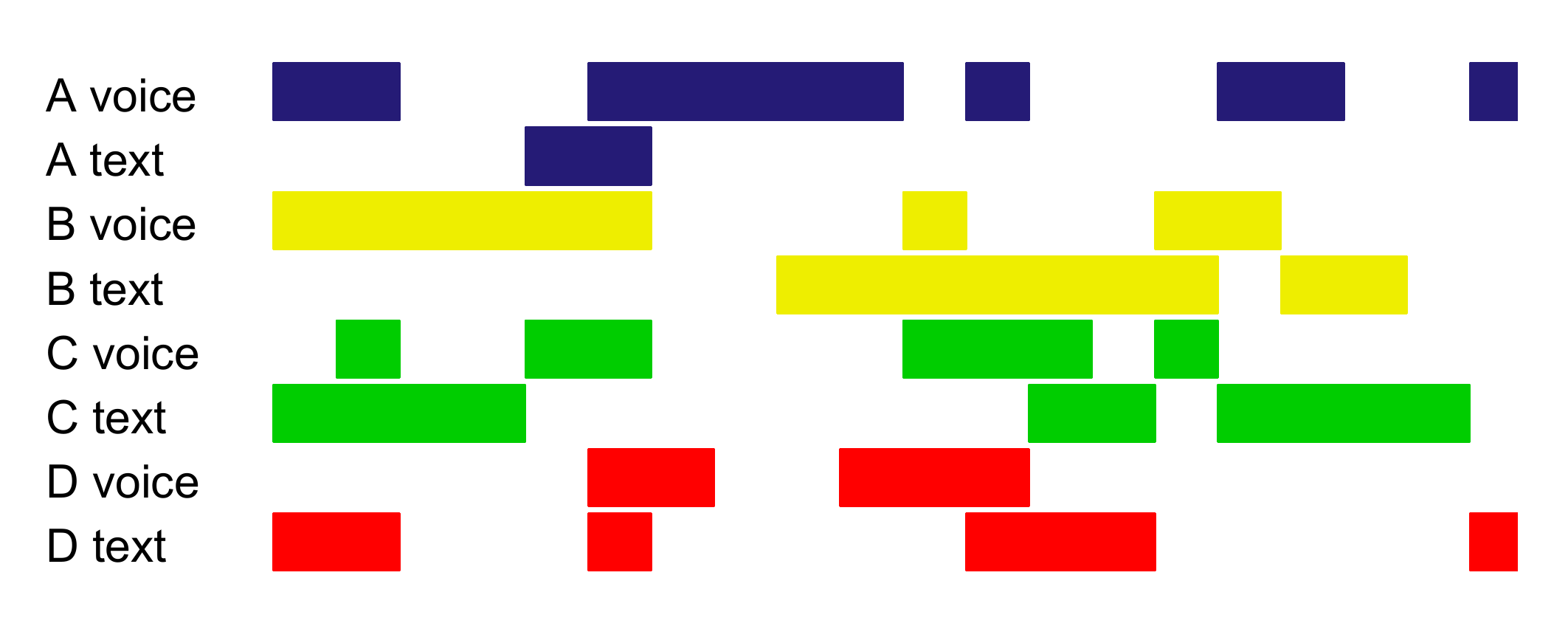}~a)
\includegraphics[width=0.45\linewidth]{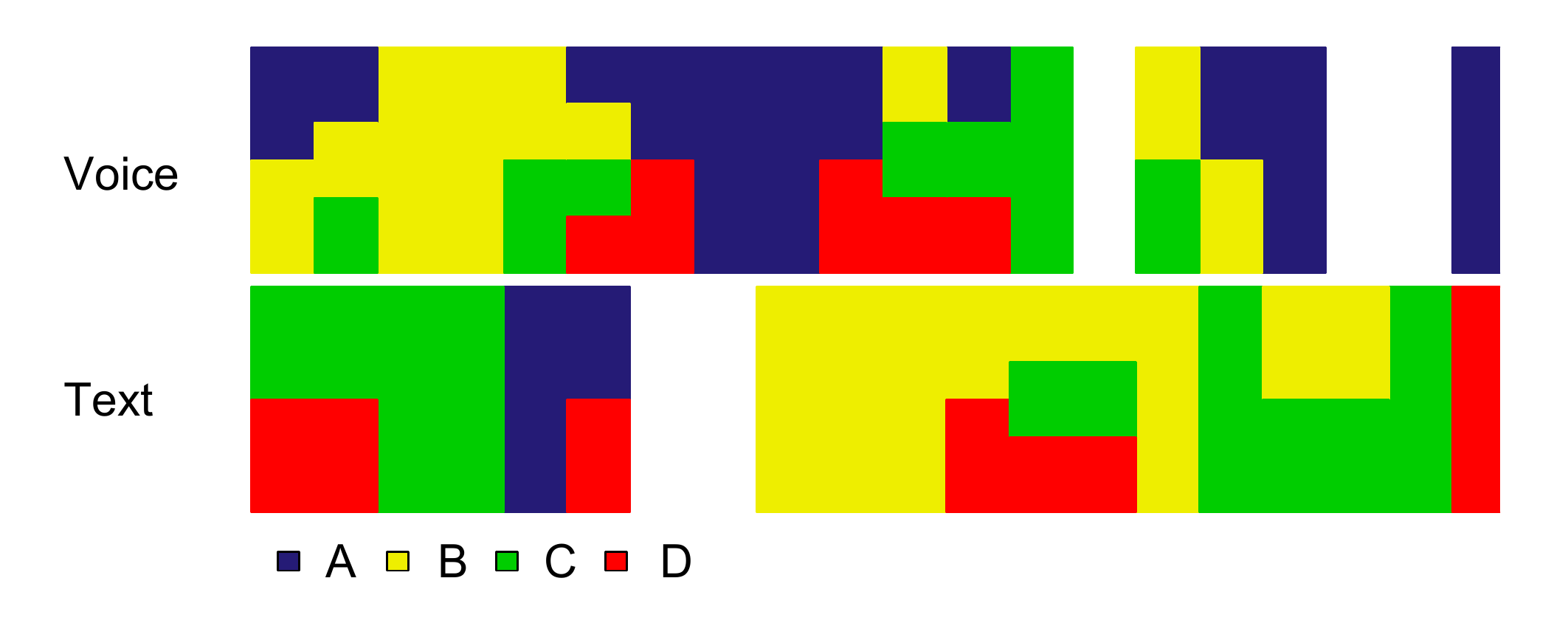}~b)
\caption{Visualization of 8 media streams (4 voice and 4 text) using a) timeline,
and b) mosaic diagrams (from \citep{bib:LuzMasoodian2007}).}
\label{fig:timeline-mosaic-diagrams}
\end{figure}

Mosaic diagrams were originally proposed by Luz and Masoodian
\cite{bib:LuzMasoodianAVI04,bib:LuzMasoodian2007}
as an alternative to conventional
timelines for visualization of temporal streams of media --- in their
case, recorded during multimedia meetings. As shown in
Figure~\ref{fig:timeline-mosaic-diagrams}, unlike timeline
visualization which reserves horizontal rows for each data stream
(Figure~\ref{fig:timeline-mosaic-diagrams}a), the mosaic visualization
uses a pre-specified vertical space proportionally between only those
streams which occur at that specific point in time
(Figure~\ref{fig:timeline-mosaic-diagrams}b).

The mosaic visualization has also been used for representation of
event schedules\cite{bib:LuzMasoodian2011}, in a manner similar to standard Gantt charts.
A study \citep{bib:LuzMasoodian2011} comparing static Gantt
charts and mosaic diagrams has shown that mosaic diagrams match Gantt
charts, in terms of speed and accuracy, for all types of tasks
requiring detection of relationships between schedule events
(e.g. durations and overlap of events).

Due to the similarity between Gantt charts and linear diagrams, we
decided to investigate the use of mosaic diagrams as a potential alternative
to linear diagrams for visualization of set relationships.
In this form, mosaic diagrams are employed as an aggregation-based set
visualization technique. 

\begin{figure}[tbp]
\centering
a)\includegraphics[width=0.3\linewidth]{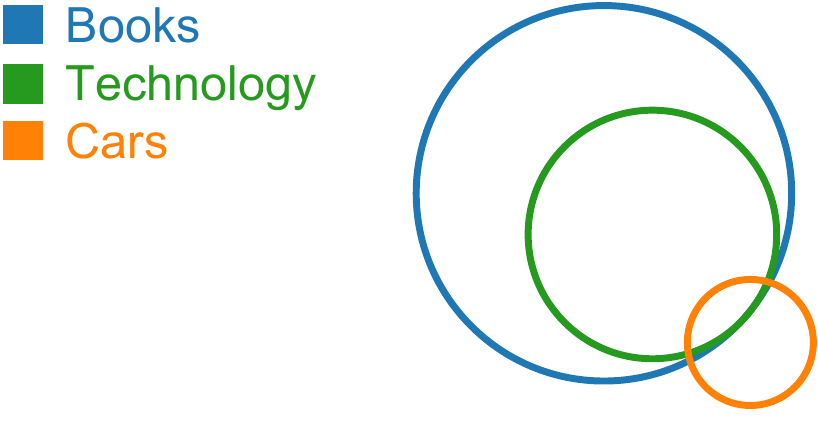}~
b)\includegraphics[width=0.3\linewidth]{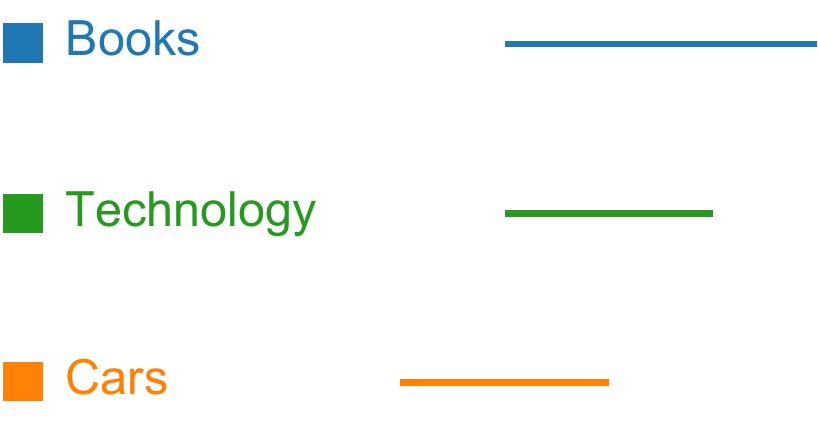}~
c)\includegraphics[width=0.3\linewidth]{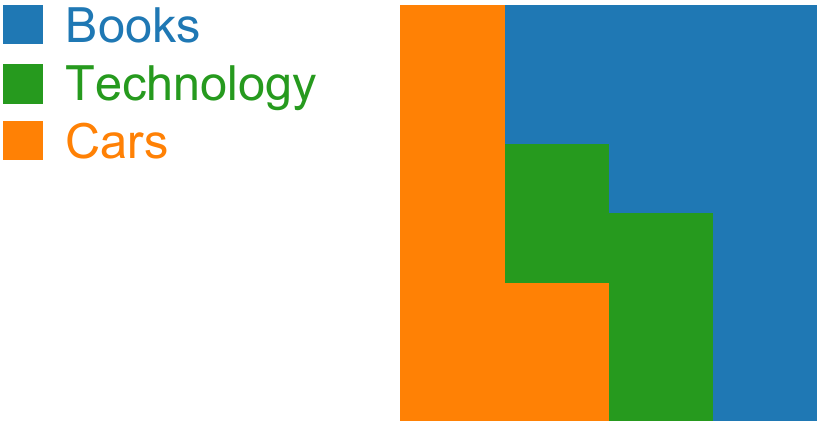}
\caption{Relationships between three example sets, shown using a) Euler, b) linear, and c) mosaic diagrams.}
\label{fig:3sets}
\end{figure}

Figure~\ref{fig:3sets} provides an example of the use of mosaic
diagrams (\ref{fig:3sets}c) to represent set relationships, in
comparison to Euler (\ref{fig:3sets}a) and linear (\ref{fig:3sets}b)
diagrams.  In this example, three sets of people are interested in
books, technology, and cars. As can be seen, some people are
interested only in books, some only in cars, some only in books and
technology, and some in all the three categories.  Furthermore,
everyone who is interested in technology is also interested in books.

\subsection{Visual Variables and Perceptual Tasks}
\label{sec:visual_variables}

Both mosaic and linear diagrams are in essence linearly-structured on
a two-dimensional plane. In terms of Bertin's graphic sign system
\citep{bib:Bertin81g,bib:Bertin67Semiologie} size and planar position
can be used to convey association. However, while for linear diagrams
these two variables would in principle suffice to communicate the
relevant set relations (intersection, disjointness and subset),
mosaics cannot avail of the alignment between horizontal bars and set
labels the way linear diagrams do. Therefore, mosaics need to employ a
further variable to distinguish the different signs for individual
sets. As there are typically many sets to label, and since colour is
generally recommended for label encoding \citep{bib:Ware12}, the
colour hue attribute was chosen as the differentiating sign in
mosaics. It should also be noted that \cite{bib:RodgersEtal2015} also
considered colour as a variable in their evaluation of linear
diagrams, but their results showed no significant differences in
performance between colour-coded and monochrome diagrams. The use of
colour places some constraints on mosaic diagrams. Notably, it limits
the number of sets that can be encoded to the number of colours that
can be reliably distinguished from each other if colour continuity
issues are to be avoided. A study by \cite{bib:Healey96c} places this
limit at 10 distinct hues. In order to maximise contrast in the mosaic
one should not choose a colour that lies in the convex hull (in a
uniform colour space) of the colours already in use. Thus a suitable
set of colours might be, for instance, the edges of a convex hull in
the CIEluv space \citep{bib:Ware12}. The use of high-saturation
colours would also help improving discrimination of mosaic areas, as
would the addition of thin, high luminance contrast boundaries to the
different tiles. As will be discussed below, in the study reported
here, we limited the use of colours to those colours used in the
experiments of \cite{bib:RodgersEtal2015} in order to reduce the
possibility of introducing confounds in the conditions we compared.

In terms of perceptual tasks, viewers rely on their ability to verify
the alignment of lines accurately in interpretation of linear and mosaics
diagrams.  As such, both types of diagrams benefit from (and to some
extent depend on) the hyperacuity characteristic of the human visual
perception \citep{bib:Westheimer09h}.  This allows viewers to perform
alignment tasks, as well as comparing length of lines, very
effectively, even in small diagrams. Unfortunately however,
performance on such tasks is known to degrade significantly if the
lines to be compared are placed too far apart in the
visual space, or when that space is crowded by 
intervening lines \citep{bib:LeviKleinAitsebaomo85v}. Furthermore, comparisons also
become more challenging in the absence of contrast between the lines
and their surrounding visual context (i.e. the background visual
space) \citep{bib:Westheimer09h,bib:SayimWestheimerHerzog08c}. These
factors have indeed contributed to, and demonstrated through empirical
studies, to suggestions made by \cite{bib:RodgersEtal2015} for generating the most effective visual
variants of linear diagrams for visualization of set relationships, as
discussed previously.

Therefore, we speculated that mosaic diagrams may be more effective
than linear diagrams for Vernier acuity tasks due to their
space-filling characteristic.  This would make visual tasks such as
identifying set relationships easier in mosaic diagrams, where
background visual space is often filled using the colour(s) associated
with set(s) of interest, unless of course when there are no
relationships between sets which is much less likely in such
visualizations. This space-filling characteristic also allows spaces
associated with sets of interest to join one another not only
horizontally, but more importantly vertically; making it easier to
perform vertical alignment tasks. It should however be pointed out
that, as is often the case in visualizations, there is a trade off in adding this space-filling
visual element. In this case, space-filling creates shapes of different
colours, which in turn can reduce detection of continuity of lines.
Although continuity is important, and according to Gestalt laws
should be preserved \citep{bib:Ware12}, mosaic relies on another
powerful Gestalt principle, namely closure \citep{bib:Ware12}, 
to allow easier detection of individual sets by creating uniquely coloured
shapes for each set.

Finally, as a side note, it should be mentioned here that although
another aggregation-based set visualization technique, called Mosaic
Plots, has previously been
proposed \citep{bib:HartiganAndKleiner1981,bib:Hofmann2000}, this
technique is rather different from the use of mosaic diagrams as
demonstrated here.  Mosaic Plots are a combination of Spine Plots and bar
charts, designed to allow representation of relationships between
groups of sets --- e.g. two gender sets, and five age group sets for
accident victims, as discussed by \cite{bib:Hofmann2000} ---
rather than direct representations of relationships between individual
sets as is the case of the mosaic diagrams investigated here.

\section{Evaluation}

In order to compare the effectiveness of mosaic and linear diagrams
for visualization of set relationships, we adopted the same set of
tasks used on the multiple comparisons of linear diagram variants
carried out by \cite{bib:RodgersEtal2015}. As in that study, the
diagrams used in our study were derived from the Twitter graph dataset
available through the SNAP project \citep{bib:LeskovecKrevl14sd}. The
variant of linear diagrams used in our comparisons was the variant
found  to be the most effective \citep{bib:RodgersEtal2015}. This variant
uses: a) \emph{heuristically minimized number of segments}, 
and b) \emph{thin horizontal lines} for representing sets.
These lines are distinguished from each other though the use of
colour, and  placed on a grid of
guide lines meant to facilitate visual alignment (see the linear
diagram shown in Figure~\ref{fig:tutorial-screenshot}, for
instance\footnote{All the content used in the evaluation is
available at \url{http://removed}}). In order to standardize the
labelling in the linear diagrams with respect to mosaic
diagrams for experimental comparison, the same legends were used in
both diagram types. These legends preserve the line ordering of
the original linear diagrams. 

The mosaic diagrams that were generated each corresponded to the
linear diagram used in the final experiment of \cite{bib:RodgersEtal2015}, except that we standardized the number
of sets to six in all tasks. We replicated the linear diagrams
manually, and used a version of the freely-available Chronos software
\citep{bib:LuzMasoodian2011}
to produce the corresponding mosaic diagrams. All images were produced
in PNG format, using the same size, colour combination, and resolution
used by Rodgers et al.  for their linear diagrams. Identical settings
were employed in the production of the corresponding mosaic diagrams.

\subsection{Methodology}
\label{sec:methodology}

Unlike \cite{bib:RodgersEtal2015}, who employed a
between-subject design and collected their data through
crowd-sourcing, we used a within-subject design, administered through
a bespoke Java application,
and recruited our participants locally by personal invitation
in each of our respective universities. 

This alternative experimental set-up was adopted in order to enable us
to recruit a smaller number of more suitable participants, and
exercise better validation and control over experimental conditions
and measurements. The choice of a within-subject (repeated measures)
design was made because it allows each participant to experience each
of the alternative visualizations under test (i.e. mosaic and linear)
repeatedly, thus mitigating the effects of any potential
inter-participant variations, and allows a smaller number of
participants usually to reveal the relevant differences, should 
such differences exist. Well known shortcomings of this kind of repeated
measures design were also addressed. Specifically carry-over effects
were mitigated by alternation of the two conditions, as well as
replications with the opposite alternation ordering (see
Table~\ref{tab:questions-sets}), and practice effects were accounted
for by the ordering of tasks from easy to difficult, again in
alternation.

Furthermore, the use of a specially designed application for the study
enabled us to obtain precise answer timings, as well as collecting
subjective task difficulty ratings. Answer time and ratings allowed us
to compare the alternative visualizations in more detail, for instance
in terms of the difficulties perceived by participants when performing
similar tasks using each of the visualizations. This is in addition to
the measures used by Rodgers et al.

In this experiment we considered three factors, with the following possible 
levels:
\begin{itemize}
\item 2 visualization types: (L)inear vs. (M)osaic
\item 3 task types: (I)ntersection, (S)ubset, and (D)isjunction
\item 2 levels of difficulty: 
  \begin{itemize}
  \item (E)asy: where the task involves identifying subsets, sets that
    intersect with, or sets that are disjoint from a set $X$,
  \item (H)ard: where the task is to identify subsets or sets that
    intersect with $X \cup Y$, or sets that are disjoint from
    $X \cap Y$.
  \end{itemize}
\end{itemize}
In order to make our study comparable to that of Rodgers et al.,
we adopted the same combinations used by them for two of these factors, 
namely task types and difficulty levels.

Each participant was requested to answer 12 ($2 \times 3 \times 2$)
task questions: 6 questions against different mosaic diagrams (MEI, MES,
MED, MHI, MHS, MHD), and 6 questions against different linear diagrams
(LEI, LES, LED, LHI, LHS, LHD). 

Each diagram used in the study depicted a collection of 6 sets and
their relationships. Each question referred to a different collection
of sets.  These 6-set collections were drawn from a larger collection
of 24 possible sets. The number of
pairwise set relations (intersections, disjointness and subsets) for all
sets used in this experiment, along with their respective mosaic and
linear diagrams are shown in Table~\ref{tab:task-questions-images}. On
average, taken in pairs, these sets contain 8.4 (SD=3.6) intersection,
6.1 (3.5) disjointness, and 1.6 (1.5) subset relations.

The numbers of elements in these sets were left unspecified, as we
were only interested in assessing abstract set relations, which are
immediately supported by linear diagrams and their mosaic equivalents.
However, see the discussion section for an example of how mosaics
could support visualisation of proportional cardinality relations
through a simple modification. Although irrelevant to this study,
exact cardinality and composition of the sets used can be retrieved
from the SNAP project website.\footnote{https://snap.stanford.edu/}


As mentioned, the task questions were presented in alternation (a
mosaic diagram following a linear diagram or vice-versa). In order to
mitigate potential order effects, we distributed the questions so that
a task was never followed by another task of the same type.
Participants were assigned automatically by the system to one of the
task question sets shown on Table~\ref{tab:questions-sets}, so as to
ensure a balanced set of answers.  Thus, for instance, on the first
series, LEI (an Easy Inclusion task, with sets represented as a Linear
diagram) is followed by a different type of task (an Easy Disjointness
task) with sets represented as a Mosaic diagram (MED). The
presentation sequences also contain no consecutive presentation of the
same type of tasks (I, D, S). As regards difficulty level, we kept a
fixed ordering whereby easier questions preceded harder questions, as
mentioned earlier.  Since this ordering is consistent across the two
visualization types (i.e. experiment conditions), task difficulty
should not affect the comparisons made between the two conditions.
The results reported later in this paper showed that our labelling of
tasks according to difficulty level conformed to the participants'
levels of performance and subjective perceptions of difficulty.

\begin{table}[!t]
  \caption{The two replications of task questions in terms of the sets and
    diagrams used in the study. L=linear diagram, M=mosaic, E=easy
    question, H=hard question, I=intersection, D=disjointness, S=subset.}
  \label{tab:questions-sets}
  \begin{tabular}{@{ }c@{\hspace{.95ex}}c@{\hspace{.95ex}}c@{\hspace{.95ex}}c@{\hspace{.95ex}}c@{\hspace{.95ex}}c@{\hspace{.95ex}}c@{\hspace{.95ex}}c@{\hspace{.95ex}}c@{\hspace{.95ex}}c@{\hspace{.95ex}}c@{\hspace{.95ex}}c@{}}
    \hline \\[-2ex] 
    \multicolumn{6}{l}{\small Task set 1:}\\
    \cline{1-3}\\[-2ex] 
    1  & 6 & 2 & 4 &  3 & 5 & 7 & 12 & 8 & 10 & 9 & 11 \\   
    LEI & MED & LES & MEI & LED & MES & LHI & MHD & LHS & MHI & LHD & MHS\\
    \hline\\[-2ex]
    \multicolumn{6}{l}{\small Task set 2:}\\
    \cline{1-3}\\[-2ex] 
       1  & 6 & 2 & 4 &  3 & 5 & 7 & 12 & 8 & 10 & 9 & 11 \\  
        MEI & LED & MES & LEI & MED & LES & MHI & LHD & MHS & LHI & MHD & LHS\\  
        \hline
      \end{tabular}
      
\end{table}

The participants were instructed to answer the questions as accurately and
as quickly as possible. We measured time (T) and accuracy (A) as
the main dependent variables. Once the participants answered each question, they
were presented with a task difficulty rating for that question, 
which they were asked to complete. Ratings were entered on a Likert scale,
ranging from 1 (very easy) to 7 (very difficult). Participants
were informed that the time taken to enter the ratings was not recorded
(i.e. it was not added to their answer times).

\begin{figure}[ht]
\centering
\includegraphics[width=.6\linewidth]{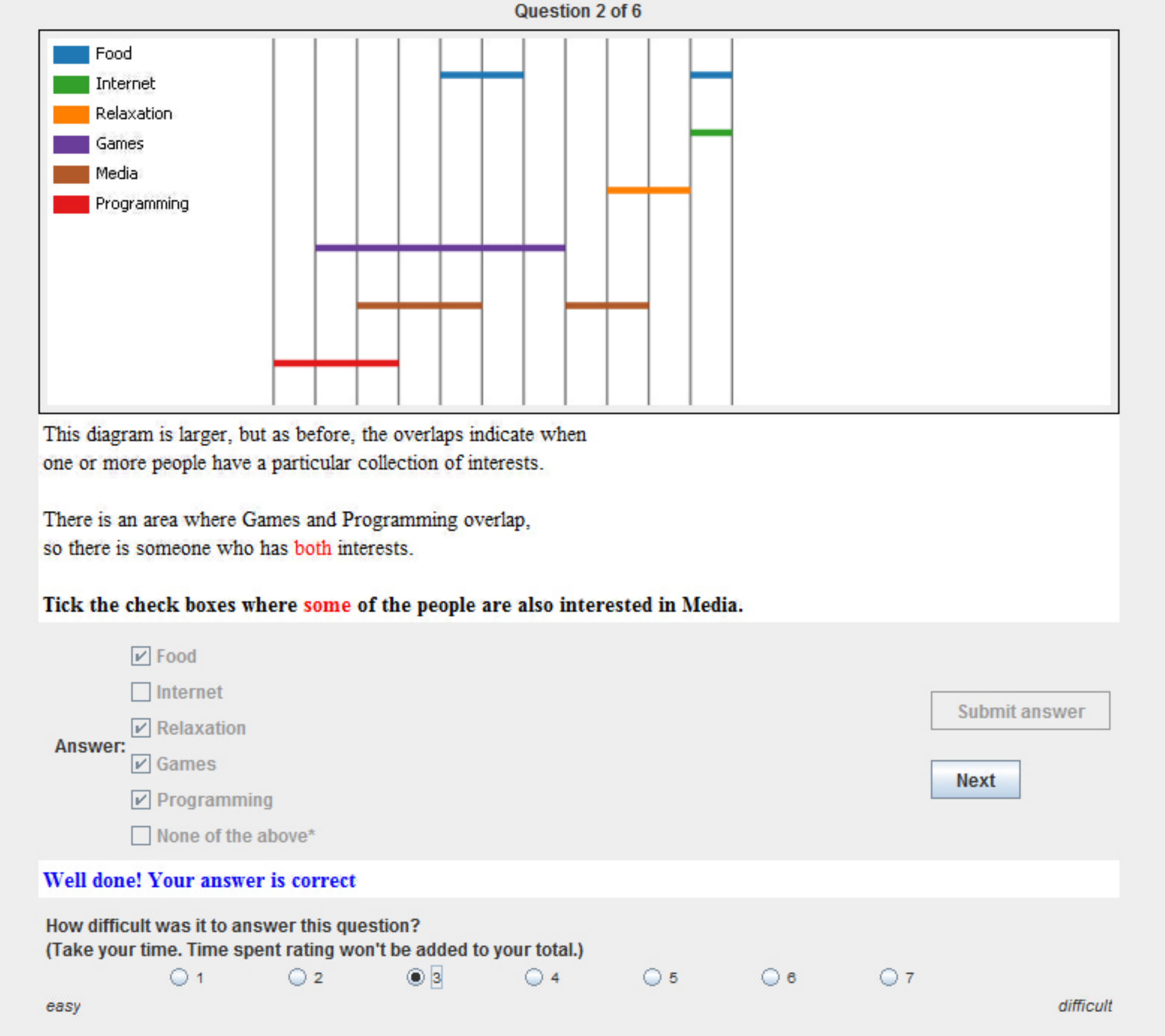}
\caption{A screen-shot of one of the tutorial questions, with the completed answer
and difficulty rating.}
\label{fig:tutorial-screenshot}
\end{figure}

A short text containing an explanation of how to interpret both mosaic
and linear diagrams, including visual examples, was presented to each
participant at the start of the study sessions. This was followed by
the participants completing a 6-question tutorial in which task
questions similar to those asked during the
actual study were presented in the same manner as in the actual study.
This tutorial set of questions was, of course, based on a different
collection of sets than the one used in the study.
After answering each of the tutorial task questions, participants were
given the correct answer, along with a brief explanation of the
answer. Figure~\ref{fig:tutorial-screenshot} shows a screen-shot of
one of the tutorial task questions, after it has been completed, along
with the difficulty rating, yet to be submitted.

After finishing the tutorial, the participants were directed to the
actual study. The study component functioned slightly differently from
the tutorial session, in that the correct answers were not presented
to the participants after they completed the test questions.

\begin{table}[h!]
  \centering
  \caption{Task questions used in the study, with all the given choices shown in brackets, and answers in \emph{italics}.}
  \label{tab:task-questions}\vspace{1ex}
  \begin{tabular}{llp{.8\linewidth}}
    \hline
      No. & Type & Question\\
    \hline
1 & EI &  Tick the check boxes where some of the people are also interested in Books.\\
& & (\emph{Android}, \emph{Cars}, \emph{Media}, \emph{News}, \emph{Stars}, None of the above) \\
2 & ES &  Tick the check boxes where all of the people are also interested in Hifi.\\
& & (\emph{Android}, \emph{Books}, Cars, \emph{Design}, \emph{Media}, None of the above)\\
3 & ED &  Tick the check boxes where none of the people are also interested in Economics.\\
& & (\emph{Cars}, Food, \emph{Music}, \emph{Stars}, \emph{Travel}, None of the above)\\
4 & EI &  Tick the check boxes where some of the people are also interested in Games.\\
& & (Computers, \emph{Design}, \emph{Food}, \emph{Programming}, Travel, None of the above)\\
5 & ES &  Tick the check boxes where all of the people are also interested in Web.\\
& & (\emph{Hifi}, iPhone, News, Relaxation, Travel, None of the above)\\
6 & ED &  Tick the check boxes where none of the people are also interested in Programming.\\
& & (Camping, \emph{Food}, \emph{Journalism}, Stars, Web, None of the above)\\
7 & HI &  Tick the check boxes where some of the people are also interested in either Computers \\
& & or Economics.
  (Games, \emph{Journalism}, \emph{News}, \emph{Relaxation}, None of the above)\\
8 & HS &  Tick the check boxes where all of the people are also interested in either Economics or Web.\\
& & (Books, Computers, Internet, Media, \emph{None of the above})\\
9 & HD &  Tick the check boxes where none of the people are also interested in both Cars and Travel.\\
& & (Design, \emph{Health}, Media, Relaxation, None of the above)\\
10 & HI &  Tick the check boxes where some of the people are also interested in either College or Relaxation.\\
& & (\emph{Android}, Design, \emph{Internet}, \emph{Stars}, None of the above)\\
11 & HS &  Tick the check boxes where all of the people are also interested in either Design or Economics.\\
& & (Food, \emph{Internet}, Relaxation, \emph{Technology}, None of the above)\\
12 & HD &  Tick the check boxes where none of the people are also interested in both Books and Food.\\
& & (Camping, \emph{Economics}, Hifi, News, None of the above)\\
    \hline
  \end{tabular}
\end{table}

\begin{table}[h!t]
  \centering
  \caption{Alternative linear and mosaic visualization images used for
    each task question. The numbers of non-empty, pairwise
    intersection (I), disjointness
    (D) and subset relations (S) are shown on the right.}
  \label{tab:task-questions-images}\vspace{1ex}
  \includegraphics[width=.8\linewidth]{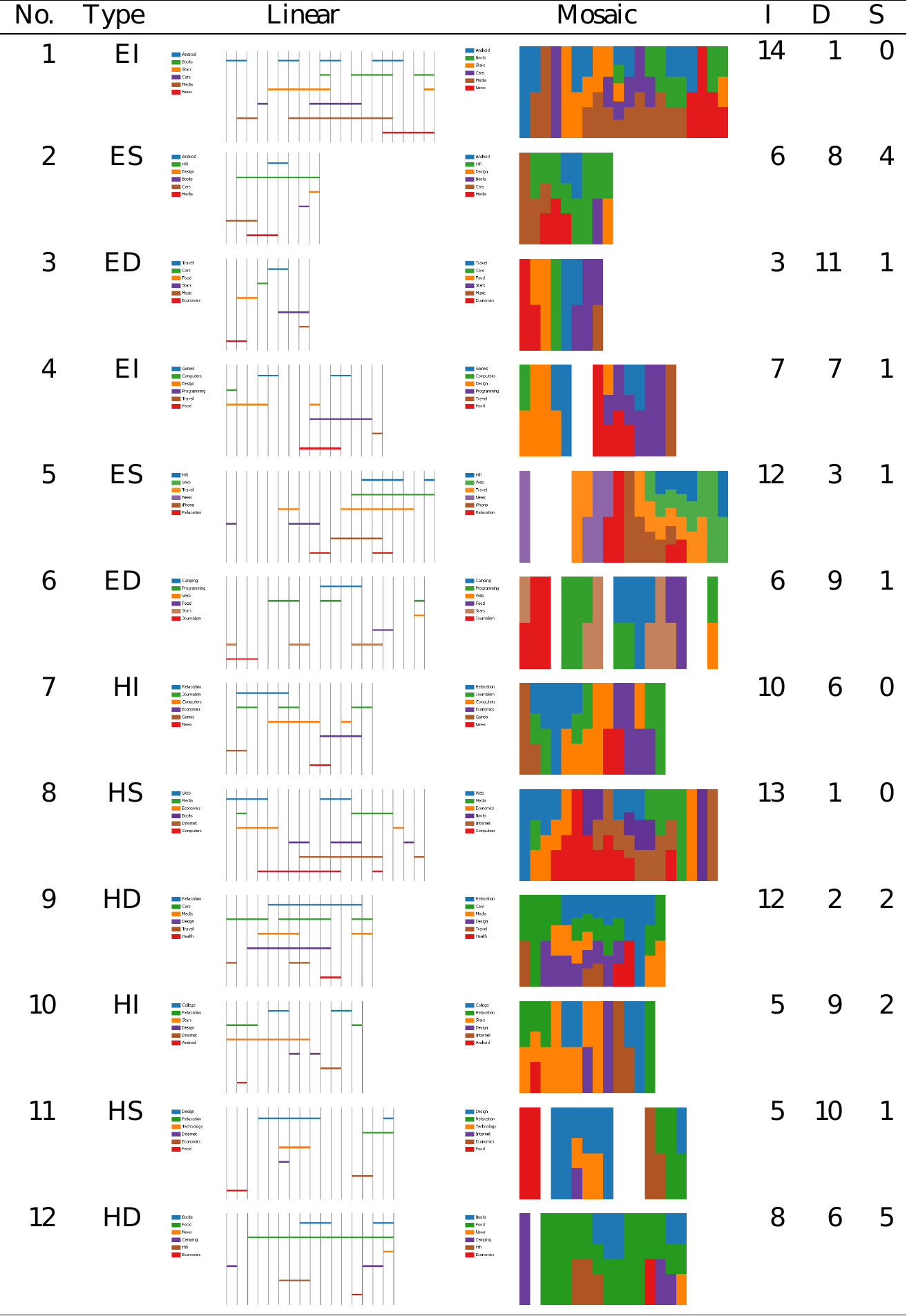}
\end{table}

\subsection{Task Questions}

Table~\ref{tab:task-questions} presents the task questions used in
this study, along with the choices given for each question (please
note that the sets belonging to the correct answers are shown in
italics).  The selected questions covered all types and difficulty
levels enumerated previously. Words representing quantifiers and
logical relations (some, all, none, both, either/or) were highlighted
in the questions, so as to draw attention to the set relations being
assessed. We realise that the wording of the questions is complicated,
and somewhat unnatural. However, given the difficulty in devising
natural-sounding questions about abstract relations, and in order to
facilitate comparison between our results and those of
\cite{bib:RodgersEtal2015}, we chose to replicate the wording used in
their experiment.

\begin{figure}[htbp]
\centering
a)\includegraphics[width=.45\linewidth]{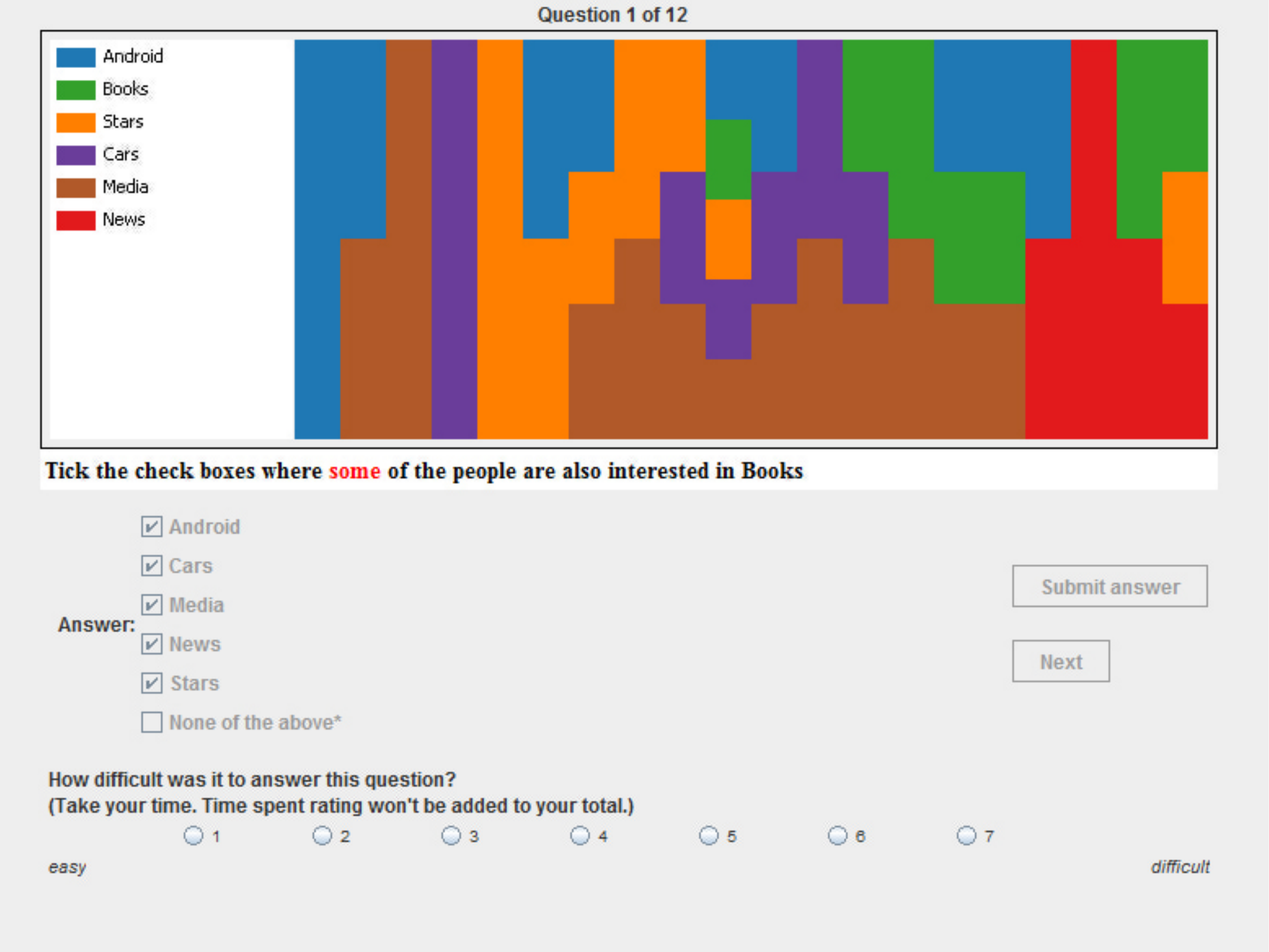}~b)\includegraphics[width=.45\linewidth]{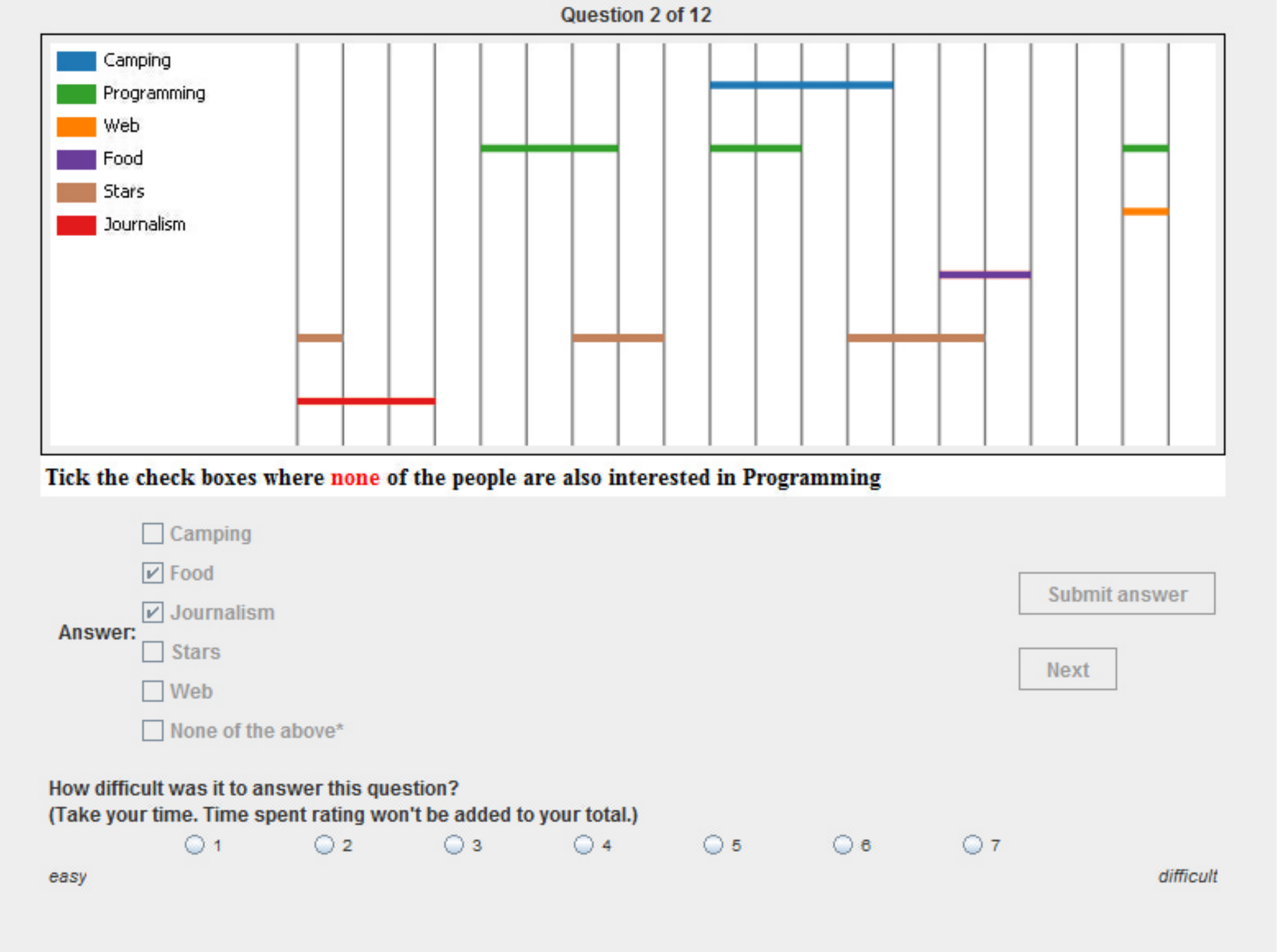}
\caption{Sample screen shots of questionnaire system, showing
  completed answer and difficulty rating: a) question 1 (MEI) and b)
  question 6 (LED), both presented during the 
experiment using Task set 2. }
\label{fig:session-screenshot-mei}
\end{figure}

Table~\ref{tab:task-questions-images} provides a small version of the 
Linear and Mosaic visualization images which were used alternatively for each question,
and of course were counter-balanced.
Figure~\ref{fig:session-screenshot-mei}.a shows a screen-shot of the Easy Intersection
question presented using the Mosaic visualization (i.e. MEI) during the
actual study session using Task set 2 (see Table~\ref{tab:questions-sets}).
Figure~\ref{fig:session-screenshot-mei}.b, on the other hand, shows
a screen-shot of the Easy Disjointness question presented using the 
Linear visualization (i.e. LED), also using Task set 2.

\subsection{Participants}

We initially conducted a power analysis to determine the number of
participants needed in order to detect differences in user performance
at the significance level $p < 0.05$. Assuming that interesting
performance differences induced by the use of mosaic or linear
diagrams would have relatively large effect sizes, say, $\eta^2$
slightly above 0.138 \citep{bib:Cohen88spanb} and aiming for 70\% power
($1-\beta$), we estimated that around 18 participants would be sufficient
for this study. 

However, we recruited 26 participants in order ensure the availability of
sufficient data. Two of these participants experienced
technical difficulties during the experiment, and their answers
were excluded from the analysis. This left us with a total of 24
participants who completed all the task questions. 
Of these, 18 were male and 6
female, and their age groups were distributed as follows: 20-29
(8), 30-39 (6), 40-49 (5), 50-59 (5). As regards their occupations, 10 were
academics, 9 students, and 5 had other occupations. Ten participants
(41.6\%) wore glasses, and none of the participants were colour blind.
Once again, due to within-subject design of our study, these variations
in participants attributes are likely to have little impact on the results of our
study.

\subsection{Results}

The answers to the task questions were collated into a single data file containing all the
288 (24x12) answers, and analysed using the R language. 

We started by comparing the accuracy scores of mosaic and linear
diagrams overall, and followed this up by comparing them according to task
type (i.e. tasks involving visual detection of intersections,
disjointness, and subsets, respectively). Analysis of accuracy figures
are of special interest here, since accuracy analysis formed the basis for performance
comparison in similar experiments \citep{bib:RodgersEtal2015}.

Pearson's $\chi^2$ test revealed no differences in either overall or
task specific comparisons. The results are summarized in
Table~\ref{tab:accuracy}. Remarkably, the overall accuracy for mosaic
diagrams was almost exactly the same as the accuracy for linear
diagrams. When broken down by task types, we see a trend (but no
statistical significance at $p<0.05$) for better
performance of 
mosaic on tasks based on the detection of intersections (questions
labelled {\em EI} and {\em HI} in tables~\ref{tab:task-questions} and \ref{tab:task-questions-images}), no
difference on disjointness tasks (questions {\em ED} and {\em HD} in
tables~\ref{tab:task-questions} and \ref{tab:task-questions-images}), and an advantage for linear diagrams
in detection of subsets (questions {\em ES} and {\em HS} in
tables~\ref{tab:task-questions} and \ref{tab:task-questions-images}). 

\begin{table}[htb]
  \centering
  \caption{Comparison of accuracy scores in task questions based on linear and mosaic diagrams. The figures represent the percentage of correct answers out of the total number of answers given.}\vspace{1ex}
  \label{tab:accuracy}
  \begin{tabular}{lccccc}
    \hline
    Task         & Linear & Mosaic &  $\chi^2$ &   $p<$ & df \\
    \hline
    Intersection & 70.8\% & 73.0\% & 0.00 & 1.00 & 1\\
    Disjointness & 77.0\% & 77.0\% & 0.00 & 1.00 & 1\\
    Subset & 79.1\% & 75.0\% & 0.05 & 0.80 & 1\\
    All & 75.6\% & 75.0\% & 0.00 & 1.00 & 1\\
  \hline
  \end{tabular}
\end{table}

Given these results, we further investigated accuracy by comparing
the different types of tasks grouped according to their difficulty levels,
that is, \emph{easy} ({\em EI}, {\em ED} and {\em ES}) versus \emph{hard}  ({\em HI},
{\em HD} and {\em HS}). In these comparisons, we employed McNemar's
test, as each group consisted of paired data. Once again the accuracy
scores were rather similar, with no statistically significant
differences shown (see Table~\ref{tab:accuracybytype}).  However,
there appears to be a tendency for greater accuracy on the easier tasks
for linear diagrams (84.7\% versus 77.8\%, $p<0.40$),  and conversely
greater accuracy for mosaic on harder tasks  (72.3\% versus 66.7\%,
$p<0.47$).

\begin{table}[ht]
  \centering
  \caption{Comparison of accuracy scores in task questions (Intersection, 
  Disjointness, Subset) grouped according to difficulty
    level (Easy, Hard) for mosaic and linear diagrams. The
  $\chi^2$ values are computed according to McNemar's method.}
  \label{tab:accuracybytype}\vspace{1ex}
  \begin{tabular}{lccccc}
    \hline
    Question         & Linear & Mosaic &  $\chi^2$ &   $p<$ & df \\
    \hline
EI & 79.2\% & 79.2\% & 0.00 & 1.00 & 1\\
ED & 87.5\% & 79.2\% & 0.12 & 0.72 & 1\\
ES & 87.5\% & 75.0\% & 0.57 & 0.45 & 1\\
HI & 62.5\% & 66.7\% & 0.00 & 1.00 & 1\\
HD & 66.7\% & 75.0\% & 0.17 & 0.68 & 1\\
HS & 70.8\% & 75.0\% & 0.00 & 1.00 & 1\\
\hline
  \end{tabular}
\end{table}

\begin{figure}
  \centering
\begin{minipage}[b]{.472\linewidth}
  \centering
  \includegraphics[width=.8\linewidth]{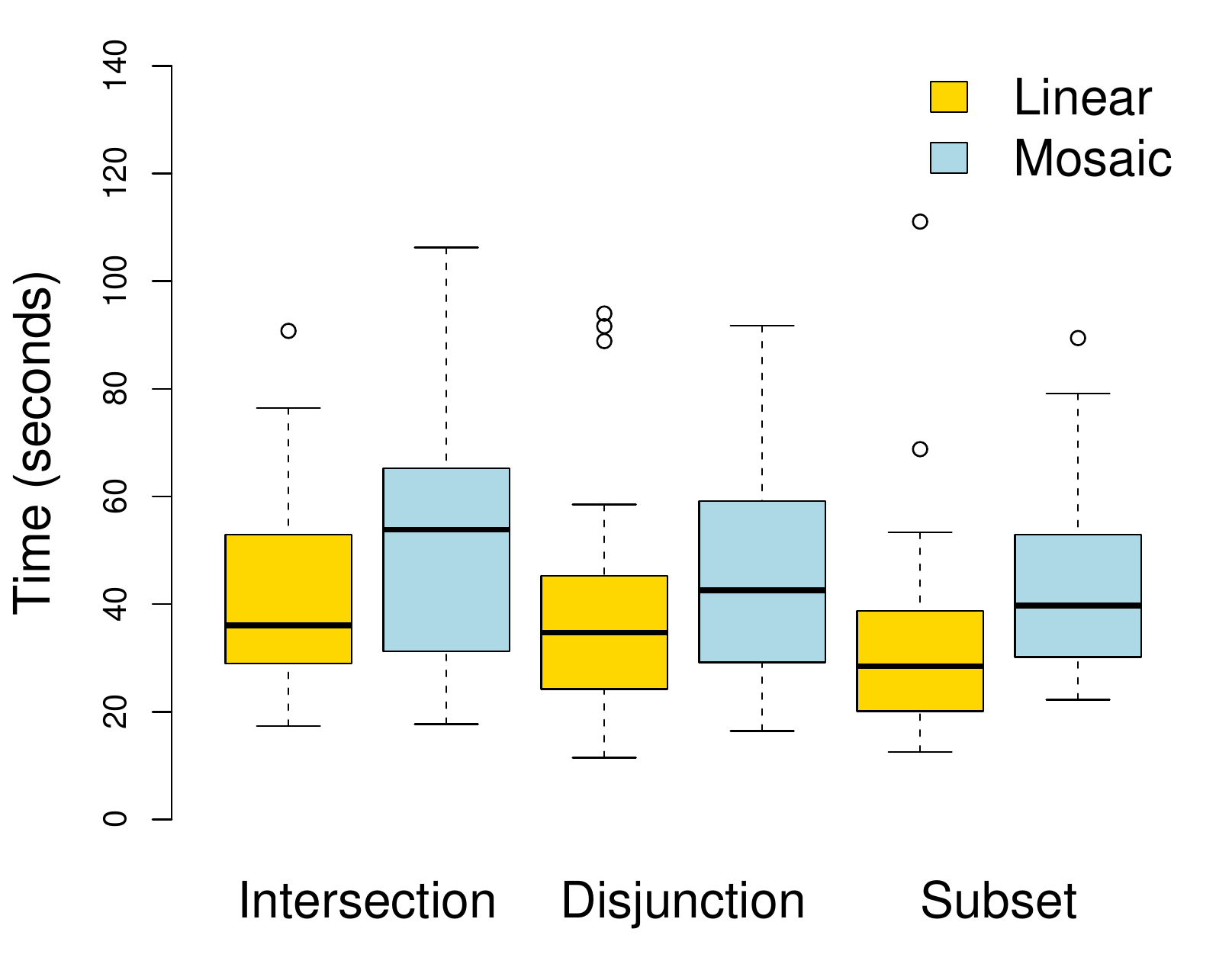}
  \caption{Time to answer {\em easy} questions using linear and mosaic diagrams.}
  \label{fig:times-easy}
\end{minipage}\hspace{2.2em}\begin{minipage}[b]{.472\linewidth}
  \centering
  \includegraphics[width=.8\linewidth]{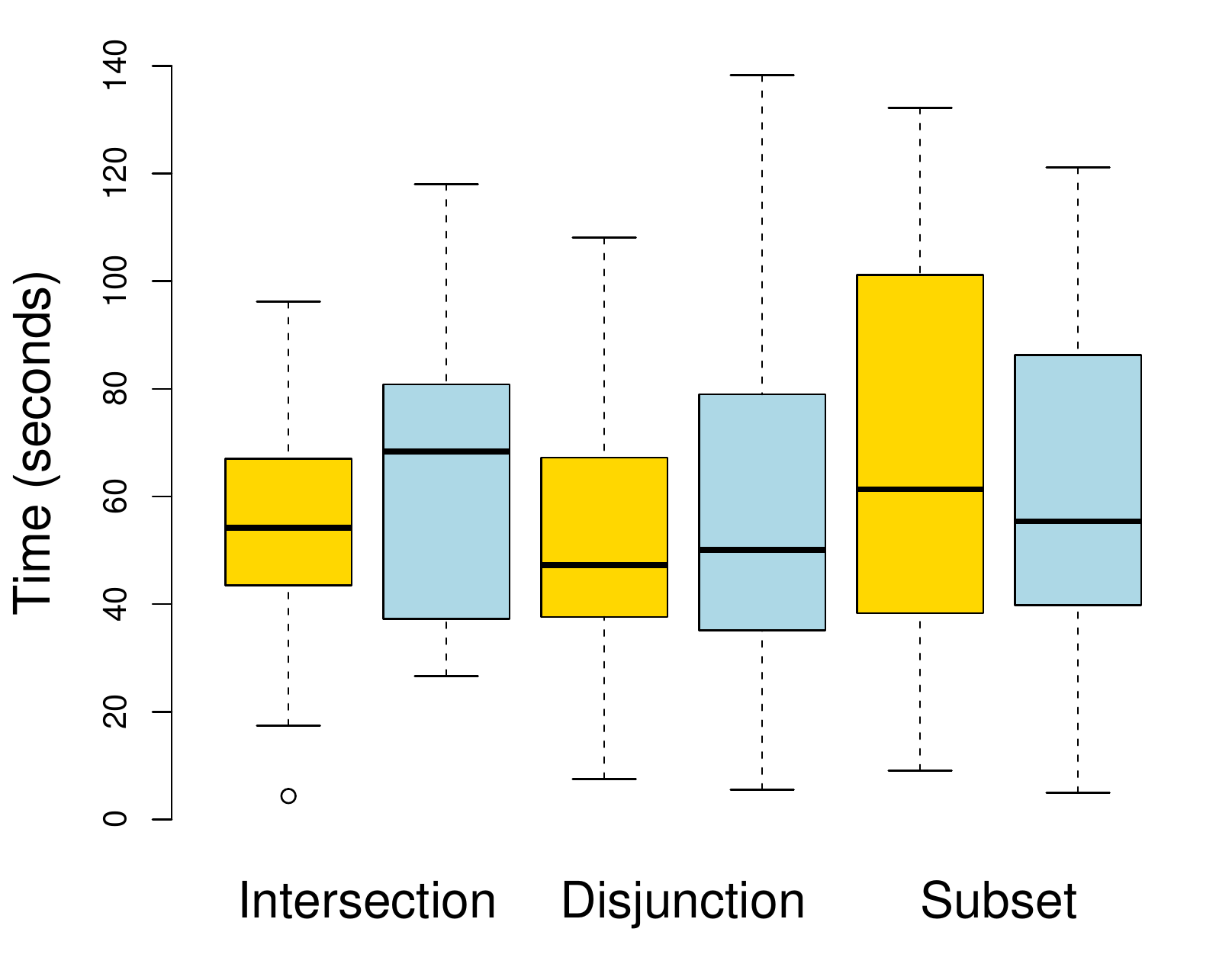}
  \caption{Time to answer  {\em hard}  questions using linear and mosaic diagrams.}
  \label{fig:times-hard}
\end{minipage}
\end{figure}

We then measured the participants' performance in terms of the time
taken to answer each task question (excluding the time taken to rate
task difficulty). The distributions of answer times are summarized on the box
plots of figures~\ref{fig:times-easy} and \ref{fig:times-hard}, for
\emph{easy} and \emph{hard} questions respectively.  Overall mosaic
users took on average 54s ($SD=27.2$)
to answer a question, while linear diagram users took 49s ($SD=27.2$).

Repeated measures analysis of variance (ANOVA) showed no
significant effects for the two visualization types ($F(1,276)=3.3$, $p=0.07$) or task
question types ($F(2,276)=0.33$, $p=0.72$). No significant interactions between
these variables were found either. 

The
only significant difference found was between \emph{easy} and \emph{hard}
tasks ($F(1,276)=31.9$, $p<0.05$, adjusted), which simply validated our
experiment design choices for task question
difficulties. Nevertheless, 
Figure~\ref{fig:times-easy}
shows a trend for users of linear diagrams to take slightly less time on the
\emph{easy} tasks. This difference does not persist however in the \emph{hard} tasks
(Figure~\ref{fig:times-hard}), reversing, in fact, for the subset type
tasks (last tasks). While further investigation is necessary to clarify this
reversal in performance, we hypothesize that it is due to the fact that
at the beginning of the experiment linear diagrams are likely to be
more familiar to users (perhaps as a consequence of  previous exposure to similar
diagrams, such as Gantt charts) than mosaic diagrams. As users gain
familiarity with the mosaic representation, their performance
improves.

\begin{figure}[ht]
  \centering
  \includegraphics[width=.55\linewidth]{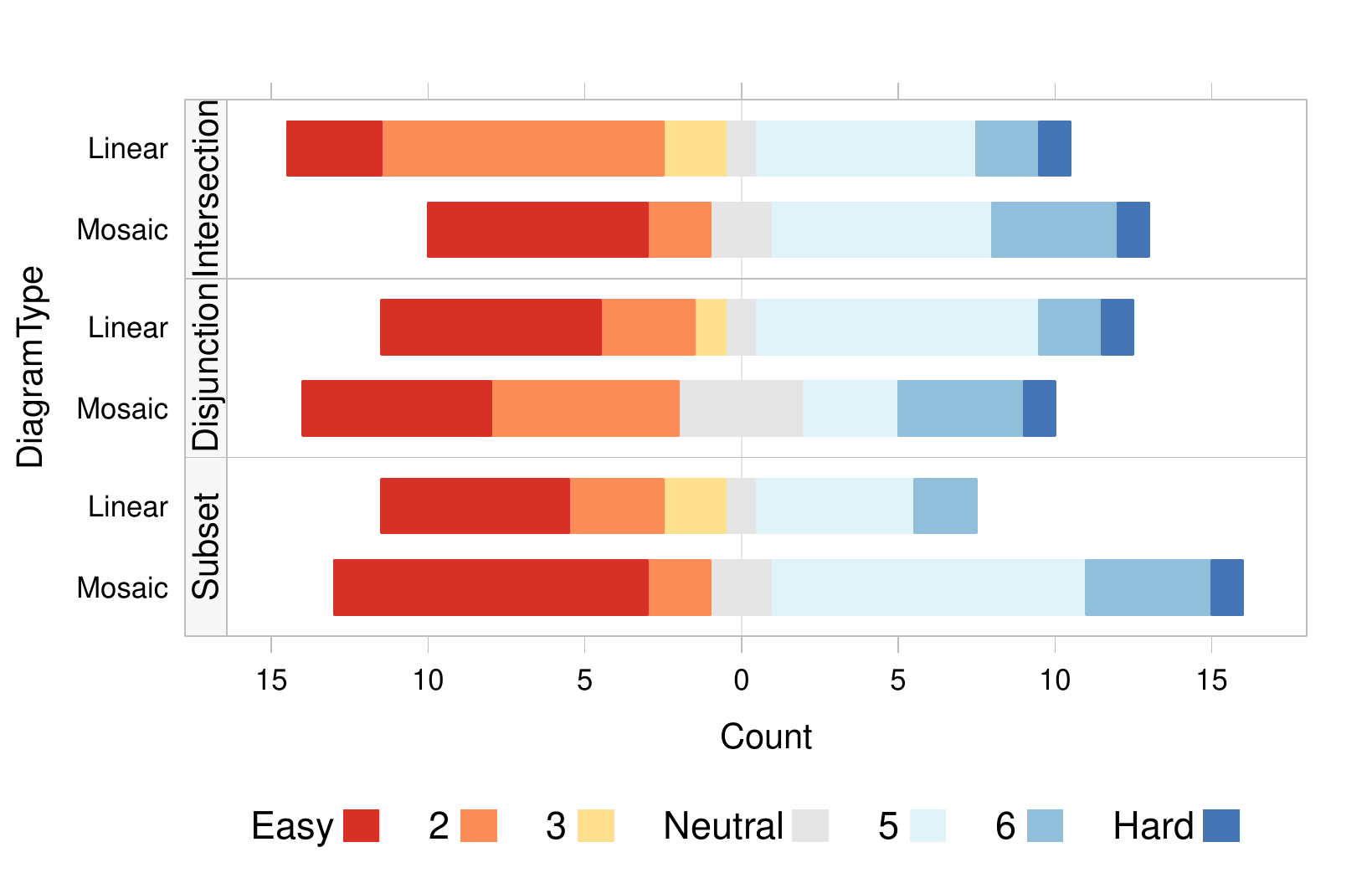}
  \caption{Ratings for task difficult, with respect to {\em easy} tasks
    ({\em EI, ED, ES}).}
  \label{fig:likert-easy}
\end{figure}

\begin{figure}[ht]
 \centering
  \includegraphics[width=.55\linewidth]{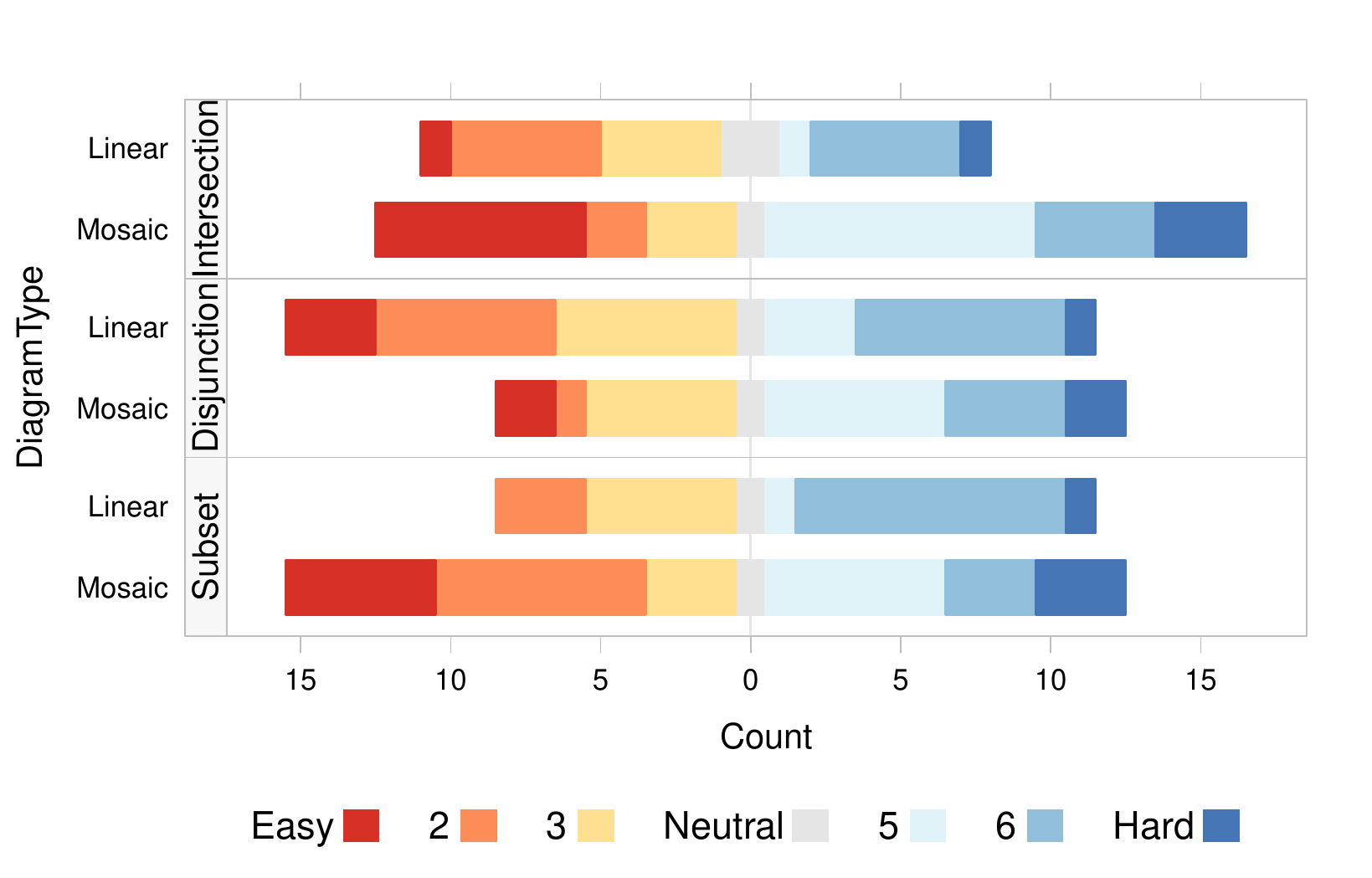}
  \caption{Ratings for task difficult, with respect to  {\em hard} tasks
    ({\em HI, HD, HS}).}
  \label{fig:likert-hard}
\end{figure}

Finally, we compared the participants' subjective ratings for task
difficulty. Figures~\ref{fig:likert-easy} and \ref{fig:likert-hard}
show summaries of responses for the two difficulty levels (\emph{easy} and \emph{hard}
respectively), grouped by the three task types and two diagram types. The
ratings are again similar, but less consistent. The median
rating is 3 for both mosaic and linear diagrams. 
The Kruskal-Wallis test showed no statistically significant difference 
($\chi^2 = 2.68$, df = 1, $p = 0.10 $). 
Despite their subjectivity, the ratings generally correlate to time on
task (Pearson's $\rho(2.91,142)=0.24$, $p < 0.01$, for mosaic diagrams,
and $\rho(3.3,142)=0.30$, $p < 0.01$, for linear diagrams) lending
additional support to the hypothesis that performance on mosaic
diagrams tended to improve more than performance on linear diagrams
over time.


\subsection{Discussion}

The study presented here has shown that ordinary mosaic diagrams are
comparable in their effectiveness to the most effective linear
diagrams that follow previously proposed visual design principles, as
discussed earlier \citep{bib:RodgersEtal2015}. However, the superiority
of temporal mosaics over temporal linear diagrams (Gantt charts) in
the context of task schedulling \citep{bib:LuzMasoodian2011}, which we
hypothesized would translate to the set comparison tasks, was not
observed in the current study. While it is not entirely clear why
accuracy and answer times were so similar for both diagrams, one could
speculate about contributing factors. One such factor may be the kinds
of tasks the user is asked to perform in each case. Even though the
basic visual tasks are roughly similar (detection of gaps and
overlaps), in schedule visualization the user is also asked to assess
interval length and position on the timeline (start and end times),
which therefore provides a structuring element which facilitates
interpretation and might benefit mosaic, where these characteristics
are represented more prominently. The complexity and level of
abstraction of the questions asked in the present set relations task
is likely to be another contributing factor. The questions in this
task are rather more abstract, and as we have pointed out, their
textual formulation has to balance naturalness with the need to avoid
ambiguity, resulting in wordings that are sometimes rather difficult
to interpret. This is likely to have played a role in levelling down user
performance across the two conditions.

There are, however, certain advantages to mosaic diagrams, which
although not tested in the current study, are likely to positively
influence their effectiveness.  For instance, the space-filling
property of mosaic diagrams preserves the overview of overlaps and
exclusions even if the diagram is dramatically reduced in size.
Linear diagrams, on the other hand, rely on the position of labels to
identify relations (colour being, as we noted before, a redundant
attribute). As these diagrams are scaled down, the user's ability to
align vertically is greatly diminished, since the horizontally aligned
labels would be impossible to preserve in miniatures, leaving the
otherwise redundant colour attribute as the only means of identifying
individual sets. In miniature linear diagrams, as in normal-sized
ones, empty spaces will dominate the image, hindering the perception
of vertical alignment of horizontal lines. Compare, for
example, \includegraphics[height=1.7ex]{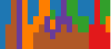}
to \includegraphics[height=1.7ex]{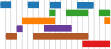}. Such miniatures
could be useful, for instance, in small-multiples
diagrams \citep{bib:Tufte90}, or in ``mini-charts'' like sparklines
\citep{bib:Tuft01} when presented along with tabular data.

In addition, mosaics highlight overlaps by facilitating visual
alignment tasks, because the edges of adjacent areas to be aligned stand out
clearly. In linear diagrams, on the other hand, comparisons of set
relationships can become increasingly more challenging as more sets
are included, thus leading to increasing vertical distances between
sets and including more distracting line segments between sets that
are placed vertically far apart. As mentioned previously, this kind of
line ``crowding'' is known to impair user performance in alignment
tasks \citep{bib:SayimWestheimerHerzog08c,bib:LeviKleinAitsebaomo85v}.

Furthermore, linear diagrams do not generally represent other set
properties such as their cardinality, and while it has been
suggested \citep{bib:RodgersEtal2015} that visual properties including
line size (e.g. length or width), colour, and texture could be used to
show set cardinality, it is acknowledged that their effectiveness has
not been demonstrated. It could be argued that the use of line length
for representing cardinality is potentially feasible, while changing
line width may be less effective, given that it has been shown that
thin lines are more effective than thick lines. Similarly, although
colour and texture visual properties can be used for representing
categorical variables, they are not very useful for representing
ordinal variables (e.g. relative cardinality of different 
sets) \citep{bib:Mackinlay86Automating}.

\begin{figure}[htbp]
\centering
a)\includegraphics[width=0.45\linewidth]{3sets-mosaic.pdf}~~
b)\includegraphics[width=0.45\linewidth]{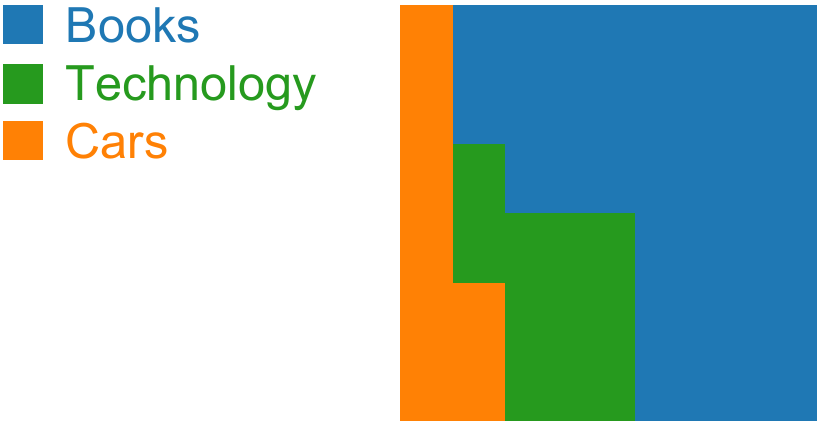}
\caption{Relationships between three example sets, shown using mosaic diagrams
a) without, and b) with cardinality comparisons.}
\label{fig:3sets-mosaic-cardinal}
\end{figure}

Mosaic diagrams, on the other hand, have been designed, and shown
\citep{bib:LuzMasoodian2011} to facilitate comparisons of relative
sizes (e.g. duration of task schedules).
Figure~\ref{fig:3sets-mosaic-cardinal} provides a simple example of
how comparison of set cardinalities could be supported by
proportionally varying the length of mosaic segments representing set
relationships in proportion to the cardinalities of the sets being
compared. Figure~\ref{fig:3sets-mosaic-cardinal}a shows only the
relationships between the three sets (Books, Technology, Cars) without
conveying any information about their cardinalities.
Figure~\ref{fig:3sets-mosaic-cardinal}b, however, makes comparisons of
the proportional cardinalities of the three sets relatively
easy. For instance, it is clear that half of the people interested in
cars are also interested in both books and technology, while the other
half are not.  Similarly, half the people interested in books are also
interested in technology, while the other half are not.  Also, it can
be seen that Books is the largest set, followed by Technology, and
Cars.

Although interactive visualization techniques are not discussed here,
mosaic diagrams have been shown to lend themselves well to the
incorporation of interactive elements (e.g. selection, brushing, zoom,
etc.) in comparison to linear-style visualizations such as Gantt
charts \citep{bib:LuzMasoodian2010}.

It should be noted, however, that as with any visualization, the use
of the colour hue attribute to encode data values places some
restrictions on the visualization for viewers who suffer from
colour-blindness. This is also true for mosaic diagrams. One possible
solution in such cases is to use another colour attribute, such as
tonal variations (i.e. value), or perhaps texture instead of hue
variations.

\section{Conclusions}
In this paper, we have proposed the use of mosaic diagrams as an
aggregation-based technique for visualization of set relationships.
This is a novel use of mosaic diagrams, which have previously been
shown to be very effective for visualization of temporal data such as
multimedia streams, and task schedules.

Although mosaics failed to yield performance improvements in
comparison to linear diagrams for set visualization tasks, as we had
expected based on reported results from a different task (schedule
visualization) which compared similar diagrams, the potential value of
mosaic diagrams for representing set relationships is supported by the
fact that mosaic produced similar results as the most effective visual
form of linear diagrams, as previously studied by \cite{bib:RodgersEtal2015}.

Finally we have discussed a number of cases where mosaic diagrams are
likely to be particularly suitable for visualization of set
relationships. These include cases where visual space is limited
and/or needs to be used more efficiently, cases where a larger number
of sets need to be represented, or cases where other set properties
such as their cardinalities also need to be presented. These, and
other interactive properties of mosaic diagrams, still need to be
further investigated within this particular task domain.  We aim to
carry out this work in the near future.


\bibliographystyle{plainnat}
\bibliography{lmsetcomptecreport}

\end{document}